# Мощные оптоэлектронные коммутаторы нано- и пикосекундного диапазона на основе высоковольтных кремниевых структур с *p-n*-переходами. I. Физика процесса переключения

А. С. Кюрегян

Всероссийский Электротехнический институт им. В. И. Ленина, 111250, Москва, Россия

Впервые проведено численное моделирование процесса переключения высоковольтных кремниевых фотодиодов, фототранзисторов и фототиристоров под действием квазиоднородного по площади освещения пикосекундными лазерными импульсами. Анализ результатов позволил получить «эмпирические» соотношения между основными параметрами коммутаторов (энергией управляющих импульсов, коэффициентом поглощения излучения, площадью структур) и параметрами, характеризующими переходной процесс переключения в цепи с активной нагрузкой. Для некоторых из этих соотношений выведены приближенные аналитические формулы, хорошо описывающие результаты моделирования. Отмечено, что различия между процессами коммутации в трех типах структур проявляются только при больших длительностях импульсов на заключительной стадии, когда восстанавливается блокирующая способность фотодиодов и фототранзисторов.

## 1. Введение

Попытки использования управляемых светом структур с *p-n*-переходами для формирования мощных высоковольтных электрических импульсов с наносекундными фронтами продолжаются уже более 40 лет [1-11]. Однако до сих пор такие коммутаторы не нашли заметного применения. Долгие время развитие этого направления импульсной техники сдерживалось низкими эксплуатационными характеристиками (энергетической эффективностью, надежностью, массогабаритными показателями, стоимостью) импульсных источников света (Nd-YAG лазеров), необходимых для управления высоковольтными кремниевыми приборами. Создание и совершенствование в течение последних 20 лет волоконных лазеров с диодной накачкой практически разрешило эту проблему[1], но для разработки высокоэффективных оптоэлектронных коммутаторов необходимо еще ясное понимание физики процесса переключения, определяющей сложную нелинейную взаимосвязь между параметрами нагрузки, полупроводниковых структур и управляющих импульсов света. Между тем опубликованные экспериментальные работы содержат лишь отрывочные сведения и имеют скорее демонстрационный, нежели исследовательский характер, а аналитические теории [13-15] не учитывают всю совокупность нелинейных эффектов, определяющих переходные характеристики структур вследствие неоднородности их легирования, зависимости подвижности от напряженности поля и ударной ионизации. В настоящей работе изложены результаты систематического исследования физических особенностей переключения таких приборов путем численного моделирования процесса переключения. В следующей статье будет рассмотрен не изучавшийся ранее вопрос об энергетической эффективности оптоэлектронных коммутаторов, которая фактически и определяет целесообразность практического применения этих приборов.

## 2. Объекты и метод исследования

Моделировался процесс переключение структур типа $p^+-p-\nu-n-n^+$ (фотодиодов), $p^+-p-\nu-n-p^+$ (фототранзисторов) и $n^{++}-p^+-p-\nu-n-p^+$ (фототиристоров) из блокирующего состояния при напряжении $U_0 = 5 \text{ kV}$ в проводящее состояние. Конструкция приборов схематично изображена на Рис. 1. Кремниевые структуры соединены с металлическими электродами. В катодном электроде имеется $m$ отверстий с площадями $S_1 \sim 1\,\text{mm}^2$, совмещенных с окнами в металлизации структур, сквозь которые проходит излучение лазера.

---

[1] Сейчас доступны коммерческие высоконадежные импульсные источники излучения с практическим КПД более 20%, длительностью импульса 0,01-10 нс, энергией 0,01-10 мДж и длиной волны около 1064 нм, почти идеально подходящей для управления высоковольтными кремниевыми приборами (например, волоконные лазеры фирмы IPG-photonics [12]).

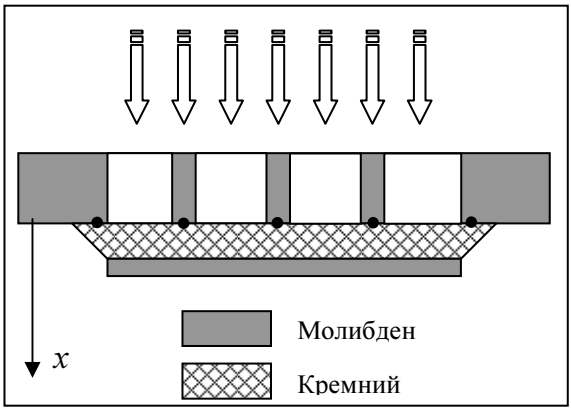

Рис. 1. Схематичное изображение поперечного сечения изучаемых полупроводниковых структур. Точками отмечено положение шунтов на поверхности катода фототиристоров.

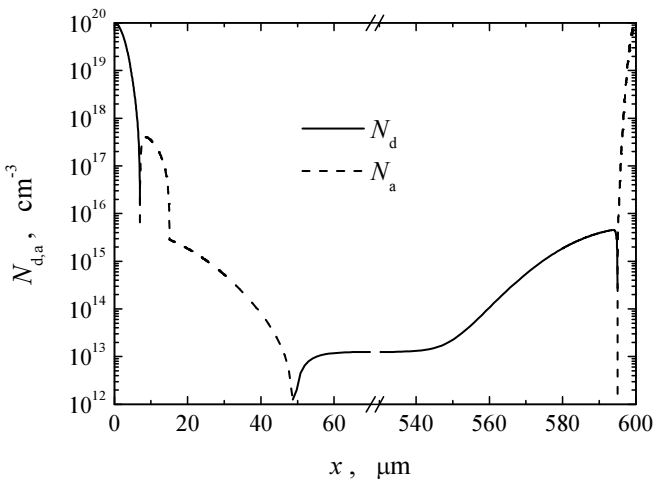

Рис. 2. Распределение доноров (сплошные линии) и акцепторов (штриховые линии) по толщине фототиристора.

Предполагалось, что полная площадь структур $S_D = 2S_{ph}$, освещаемая площадь $S_{ph} = mS_1$ изменялась в пределах $0.1 \div 3$ cm². Толщина структур $d = 600$ μm, параметры рекомбинации в модели Шокли-Рида $\tau_{n0} = \tau_{p0} = 3$ мкс, концентрация доноров в $\nu$-слое $N_0 = 1.25 \cdot 10^{13}$ cm⁻³, распределение примесей в диффузионных слоях описывалось функцией Гаусса. Использованный нами профиль легирования фототиристора $N_{d,a}(x)$, изображенный на Рис. 2, обеспечивал при наличии кольцевой шунтировки котодного $n^+ - p$ - перехода на периферии освещаемых окон и температуре 375 К напряжение пробоя $U_B = 6$ kV и скорость нарастания напряжения до 1 kV/μs. Он отличается от предложенного в [11] профиля тем, что прилегающие к $\nu$-слою диффузионные $p$-база и буферный $n$-слой примерно в 10 раз толще и в 1000 раз слабее легированы. Это сделано для уменьшения напряженности краевого поля в реальных структурах и подавления динамического лавинного пробоя на первом этапе процесса коммутации (см. далее). Структуры фототранзистора или фотодиода отличались тем, что изменялся тип проводимости одного или обоих приэлектродных сильно легированных слоев, а шунтировка отсутствовала. Их напряжения пробоя были равны 6 kV и 6.6 kV соответственно.

Мощность излучения лазера изменялась со временем по закону $P(t) = P_p F_t(t)$, где

$$F_t(t) = \sqrt{2e}\frac{t}{t_{ph}}\exp\left[-\left(\frac{t}{t_{ph}}\right)^2\right] \qquad (1)$$

где $P_p$ - пиковая мощность в момент времени $t = t_{ph}/\sqrt{2}$. При этом энергия импульса излучения $W_{ph} = \sqrt{2/e}P_p t_{ph}$. Скорость генерации электронно-дырочных пар в структуре представлялась в виде $G(x,t) = G_0 F_t(t) F_x(x)$, где $G_0 = P_p \kappa (1 - R_{ph})/\hbar \omega S_{ph}$, $\hbar \omega$ - энергия кванта, $\kappa$ - коэффициент поглощения света в полупроводнике, $R_{ph}$ - коэффициент отражения света от поверхности окна фотодиода. В настоящей работе мы считали, что на поверхность окон нанесено просветляющее покрытие и поэтому $R_{ph} = 0$. Распределение скорости генерации по толщине структур описывалось функцией

$$F_x(x) = 2e^{-\kappa d}\operatorname{ch}\left[\kappa(d-x)\right]. \qquad (2)$$

При этом мы пренебрегали относительно слабой зависимостью $\kappa$ от $x$ вследствие эффекта Франца-Келдыша и сужения запрещенной зоны при высоком уровне легирования и считали тыловой контакт (плоскость $x = d$) идеально отражающим.

Динамика электронно-дырочной плазмы в структурах моделировалась с помощью программы «Исследование» [16] при начальном напряжении на них $U_0 = 5$ kV и температуре 75 C. В настоящей работе изучался простейший режим, соответствующий разряду формирующей линии длиной $l$ с волновым сопротивлением $Z$ на согласованную активную нагрузку $R_L = Z$. В этом случае через кремниевую структуру и нагрузку протекает импульс тока с длительностью $2l/c$ и амлитудой $(U_0 - U_{\min})/R$, где $c$ - скорость света в диэлектрике, заполняющем линию, $R = 2R_L$, $U_{\min}$ - минимальное падение напряжения на структуре. Далее изложенны результаты, полученные при $t_{\text{ph}} = 10$ ps для значений $R_L = 5$ Ohm и, если это особо не оговорено, $S_{\text{ph}} = 0.5$ cm$^2$, $\kappa = 32$ cm$^{-1}$ (согласно [17] примерно такое значение $\kappa$ имеет излучение с длиной волны 1045 нм, типичной для волоконных лазеров, в Si при 75 C).

## 3. Результаты моделирования и их обсуждение

Результаты моделирования приведены на Рис. 4-12. Процесс коммутации состоит из ряда последовательных этапов (см. Рис. 3), которые сначала мы опишем для фотодиодов.

### 3.1. Переключение в проводящее состояние

На первом этапе управляющее излучение порождает наравновесные электроны и дырки с концентрациями $n(x,t), p(x,t)$ и через структуры начинает протекать ток проводимости с плотностью $j = q(nv_n + pv_p)$, где дрейфовые скорости $v_{n,p} = \overline{v}_{n,p} E / (E + \overline{E}_{n,p})$, $\overline{v}_{n,p} = \mu_{n,p} \overline{E}_{n,p}$, $\mu_{n,p}$ - низкополевые подвижности, $E$ - напряженность поля. Этот ток разряжает емкость структур $C_D = \varepsilon S_{\text{ph}}/w_0$ и индуцирует ток нагрузки (см. Рис. 4) равный [15]

$$I_L(\theta) = S_{\text{ph}} \int_0^\theta \overline{j}(\theta') \exp(\theta' - \theta) d\theta', \quad (3)$$

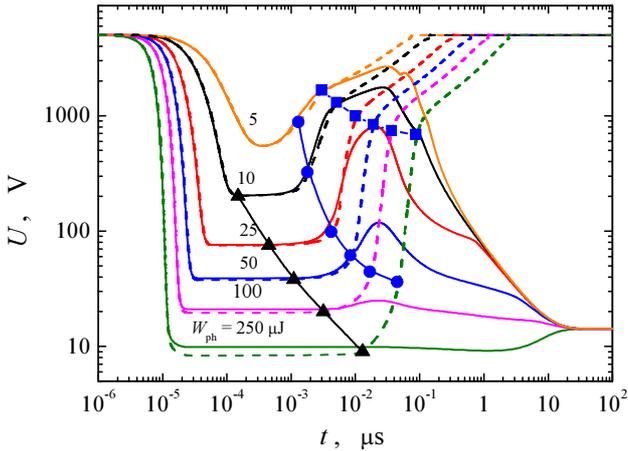

Рис. 3. Вольт-секундные характеристики процессов коммутации фотодиодов (штриховые линии) и фототиристоров (сплошные линии) при различных $W_{\text{ph}}$. Символами отмечены моменты восстановления $p-\nu$-перехода $t_{0p}$ (треугольники), начала формирования ОПЗ $t_{\text{sc}}$ (кружки) и начала динамического лавинного пробоя $t_{\text{av}}$ (квадраты) фотодиода.

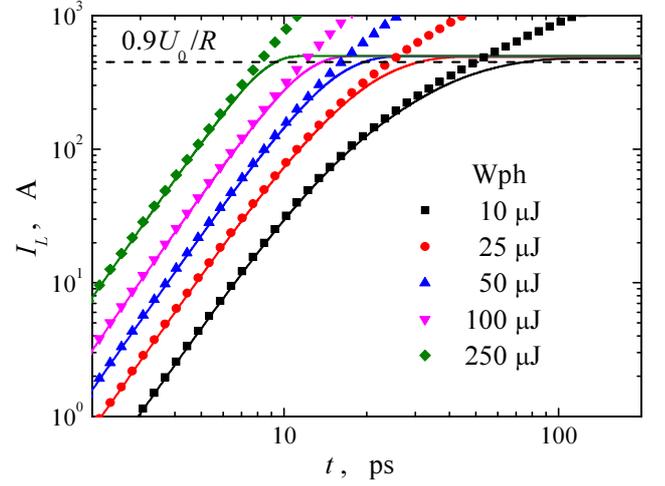

Рис. 4. Зависимости тока нагрузки от времени на первом этапе процесса коммутации фотодиода при различных энергиях импульса света $W_{\text{ph}}$. Сплошные линии - результаты численного моделирования, символы – расчет по формуле (5).

где безразмерное время $\theta = t/\tau_D$, $\tau_D = RC_D$, $\overline{j}$ - усредненная по толщине $w_0 = x_n - x_p$ истощенного слоя плотность тока электронов и дырок, $x_{p,n}$ - границы истощенной области. При этом падение напряжения на структурах $U = U_0 - I_L R$ быстро уменьшается, как изображено на Рис. 3. Простую оценку $\overline{j}$ можно сделать для малых времен, когда еще $t v_{n,p} \ll \min(d, \kappa^{-1})$, $n \approx p$, $E > \overline{E}_{n,p} \sim 10$ kV/cm почти во всем истощенном слое и поэтому

$$\overline{j}(t) \approx q(\overline{v}_n + \overline{v}_p) G_0 w_0^{-1} \int_{x_p}^{x_n} F_x(x) dx \int_0^t F_t(t') dt'. \quad (4)$$

Подстановка (4) в (3) дает после интегрирования с учетом (1),(2) следующую формулу для тока нагрузки

$$I_L(\theta) \approx I_0 \left[ 1 - e^{-\theta} \Phi(\theta, \tau_d/t_{\text{ph}}) \right], \quad (5)$$

$$I_0 = q \frac{W_{\text{ph}}}{\hbar \omega} \frac{\overline{v}_n + \overline{v}_p}{w_0} (1 - R_{\text{ph}}) 2 e^{-\kappa d} \left\{ \text{sh}\left[\kappa(d - x_p)\right] - \text{sh}\left[\kappa(d - x_n)\right] \right\},$$

$$\Phi(\theta, \gamma) = 1 + \frac{\sqrt{\pi}}{2\gamma} \exp\left(\frac{1}{4\gamma^2}\right) \left[ \text{erf}\left(\gamma\theta - \frac{1}{2\gamma}\right) + \text{erf}\left(\frac{1}{2\gamma}\right) \right]$$

Эта формула хорошо описывает зависимость $I_L(t/\tau_D)$ вплоть до значений $I_L \approx 0.7 U_0/R$ (см. Рис. 4), после чего она дает завышенные значения, так как не учитывает снижение дрейфовых скоростей вследствие уменьшения со временем напряженности поля до значений, меньших $\bar{E}_{n,p}$ (см. Рис. 5). Тем не менее, она позволяет оценить «инженерное» время коммутации $t_{0.9}$, являющееся решением уравнения $I_L(t_{0.9}/\tau_D) = 0.9U_0/R$, занижая его на 20-30%. Эту погрешность можно скомпенсировать, используя для расчета $t_{0.9}$ модифицированное уравнение

$$I_L\left(\frac{4t_{0.9}}{5\tau_D}\right) = 0.9U_0/R. \qquad (6)$$

Введение в (6) дополнительного множителя 4/5 обеспечивает описание зависимостей $t_{0.9}$ от $W_{ph}$ (см. Рис. 6), $S_{ph}$ (см. Рис. 7) и $\kappa$ (см. Рис. 8) с достаточной для практических целей точностью, если коэффициент поглощения $\kappa < 40$ cm$^{-1}$. При бо́льших $\kappa$ распределение неравновесных носителей заряда в $\nu$–слое становится очень неоднородным, условие $E > \bar{E}_{n,p}$ применимости формулы (5) нарушается гораздо раньше и поэтому уравнение (6) дает сильно заниженные значения $t_{0.9}$, как показано на Рис. 8.

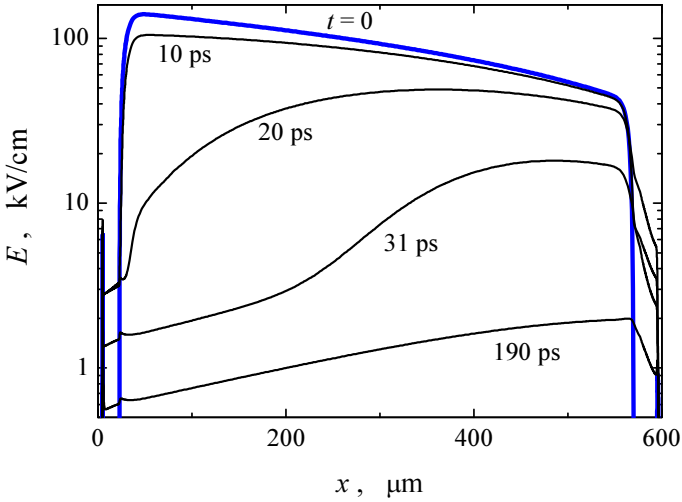

Рис. 5. Распределения электрического поля в фотодиоде при $W_{ph} = 25$ μJ в различные моменты времени; «инженерное» время коммутации $t_{0.9} = 31$ ps.

Следует отметить качественное различие распределений поля $E(x,t)$, изображенных на Рис. 5, и полученных ранее в работе [15] для $p^+ - \nu - n^+ -$структур со ступенчатым легированием (см. Рис.5 в [15]). Оно состоит в том, что наличие относительно толстых и слабо легированных диффузионных слоев подавляет образование тонких областей с большой напряженностью поля на границах $\nu-$слоя и таким образом предотвращает преждевременное наступление динамического лавинного пробоя структур.

### 3.2. Стадия высокой проводимости

Второй этап наступает при $t > t_{0.9}$, когда напряженность поля становится меньше $\bar{E}_{n,p}$, напряжение на структуре $U(t)$ быстро уменьшается и достигает своего минимума

$$U_{\min} = U_0\left[1+(1-R_{ph})\frac{W_{ph}}{\hbar\omega}\frac{q(\mu_n+\mu_p)R_L}{d^2}\Psi(\kappa d)\right]^{-1}, \quad (7)$$

где $\Psi(x) = 2x^2 e^{-x}\left[\arctg(\sh x)\right]^{-1}$, ток нагрузки увеличивается и достигает максимума $I_{\max} = (U_0 - U_{\min})/R$, а ток через структуру $jS_{ph}$, намного превосходящий $I_L$ во время первого этапа, быстро уменьшается и сравнивается с $I_L$. Формула (7) хорошо описывает зависимости $U_{\min}$ от $W_{ph}$ (см. Рис. 7) и $\kappa$ (см. Рис. 8), но не учитывает электронно-дырочное рассеяние, которое ослабевает при снижении концентрации неравновесных носителей заряда. Вследствие этого $U_{\min}$ уменьшается примерно на 20 % при увеличении $S_{ph}$ в 20 раз и $W_{ph} = const$.

### 3.3. Восстановление $p^+$-$n$- и $n$-$n^+$-переходов.

Неравновесные электроны и дырки, накопленные в структурах во время освещения, начинают вытягиваться из приграничных областей. В фотодиодах это приводит к восстановлению[2] $p^+ - p-$ и $n - n^+ -$переходов, расположенных в плоскостях $x_p^j$ и $x_n^j$ в моменты $t = t_{0p}$ и $t = t_{0n}$. Аналитическая оценка $t_{0p,0n}$ возможна в простейшем случае толстых однородно легированных $p^+ -$ и $n^+ -$слоев при низком уровне инжекции в них и высоком уровне инжекции в прилегающих $p -$ и $n -$слоях. Для времени восстановления $p^+ - p -$перехода получается

$$t_{0p} = \frac{\pi}{4}\left[\frac{\mu_n+\mu_p}{\mu_n}\frac{n_0}{j}\left(\sqrt{D}+\sqrt{D_n^+}\right)\right]^2, \qquad (8)$$

---

[2] Восстановлением этих переходов мы, как и в [18], называем процесс, в результате которого концентрация основных носителей заряда на границах $x = x_{p,n}^j$ практически сравнивается с $N_{a,d}$ соответственно.

где $n_0 = \kappa W_{ph}(1-R_{ph})/\hbar\omega S_{ph}$, $D_n^+$ - коэффициент диффузии электронов и дырок в $p^+$ – слое, $D = 2D_n D_p/(D_n + D_p)$ - амбиполярный коэффициент диффузии в слабо легированном Si. Формула (8) дает заниженное время $t_{0p}$. Однако это практически не влияет на точность расчета $t_{sc}$ в следующем разделе, так как «правильные» значения $t_{0p}$, полученные путем численного моделирования, примерно в 10 раз меньше $t_{sc}$ (см. Рис. 6).

Восстановление завершается образованием крутых концентрационных фронтов, движущихся навстречу друг другу и отделяющих центральную плазменную область $x_p^f < x < x_n^f$ от областей $x < x_p^f$, $x > x_n^f$, практически свободных от неравновесных носителей заряда [20,20].

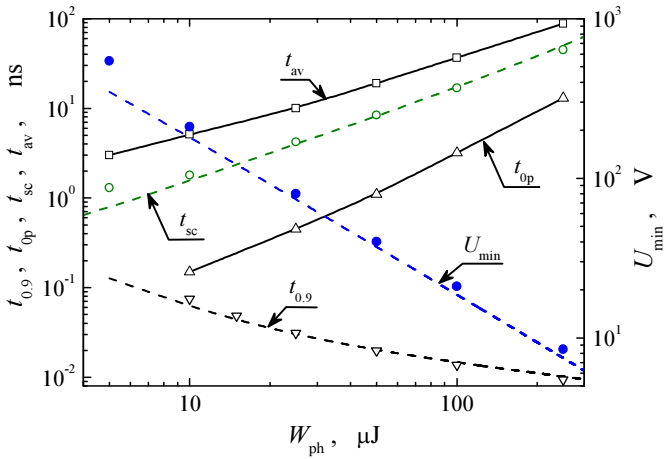

Рис. 6. Зависимости времен $t_{0.9}, t_{0p}, t_{sc}$ и напряжения $U_{min}$ (темные символы) на фотодиоде от $W_{ph}$. Символы - результаты численного моделирования, штриховые линии – расчет по формулам (6), (7) и (10).

### 3.4. Формирование области пространственного заряда (ОПЗ).

Концентрации легирующих примесей $N_{a,d}(x_{p,n}^f)$ на фронтах уменьшаются со временем. Поэтому при $t = t_{sc}$ в одной из областей $x < x_p^f$, $x > x_n^f$ нарушается условие квазинейтральности [20,20]. Если $\kappa$ не слишком велик, то сначала это происходит в области $x < x_p^f$ при $x_p^f = x_p^{sc}$, где $x_p^{sc}$ - решение уравнения [20]

$$\left|N_a(\tilde{x}_p)\right| = \frac{1}{q}\left[\frac{j}{\overline{v}_p} + \sqrt{\frac{\varepsilon j}{\gamma\mu_p\lambda_p(\tilde{x}_p)}}\right], \quad (9)$$

где $\lambda_p^{-1}(\tilde{x}_p) = \partial \ln N_a(\tilde{x}_p)/\partial x$, $\gamma \sim 0.1$ - параметр, характеризующий степень нарушения нейтральности. Численное решение (9) позволяет вычислить $t_{sc}$ по формуле [20]

$$t_{sc} \approx t_{0p} + \frac{\mu_n + \mu_p}{\mu_n}\frac{n_0}{j}(x_p^{sc} - x_p^j) \quad (10)$$

При выводе (10) мы пренебрегали слабыми зависимостями тока $S_{ph}j$ от времени при $t_{0p} < t < t_{sc}$ (см. Рис. 4)) и концентрации неравновесных электронов перед фронтом от координаты

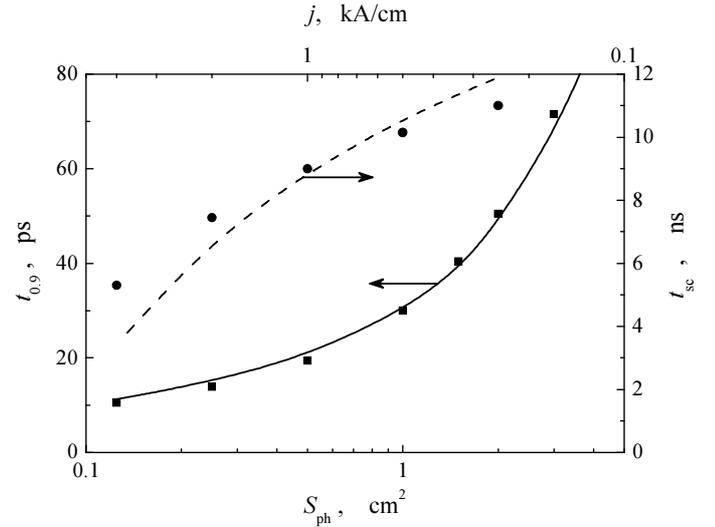

Рис. 7. Зависимости времен $t_{0.9}, t_{sc}$ фотодиодного коммутатора от $S_{ph}$ при $W_{ph} = 50$ μJ. Символы - результаты численного моделирования, линии – расчет по формулам (6) и (10).

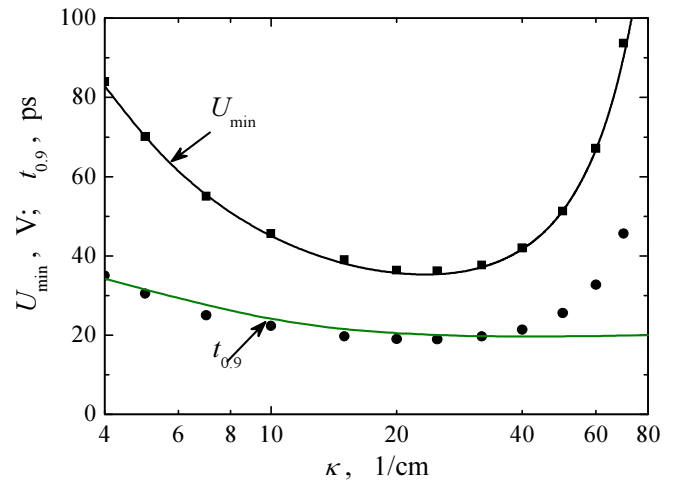

Рис. 8. Зависимости времени коммутации $t_{0.9}$ и напряжения $U_{min}$ от коэффициента поглощения $\kappa$ для фотодиода при $W_{ph} = 50$ μJ. Символы - результаты численного моделирования, линии – расчет по формулам (6) и (7).

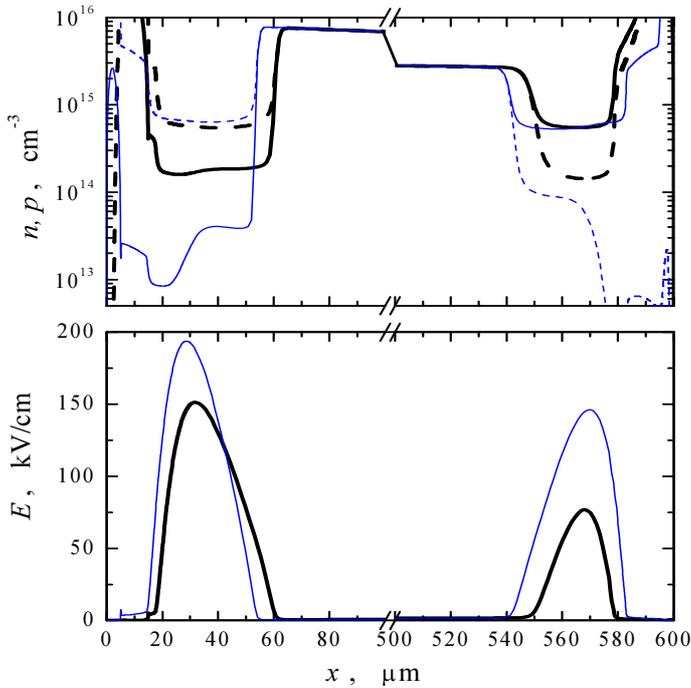

Рис. 9. Распределения электронов (*n*, сплошные линии), дырок (*p*, штриховые линии) и электрического поля (*E*) в фотодиоде (тонкие линии) и фототиристоре (толстые линии) при $W_{ph} = 25$ μJ в момент времени $t = t_{av} = 10$ ns.

(вследствие неравенства $\kappa x_p^{sc} \ll 1$), а также полагали, что $N_a \ll n_0$ в области $x_p^j < x < x_p^{sc}$. Последнее условие нарушается при малых $W_{ph}$ и больших $S_{ph}$, однако формула (10) неплохо описывает зависимости $t_{sc}$ от $W_{ph}$ (см. Рис. 6) и $S_{ph}$ (см. Рис. 7).

После нарушения нейтральности начинается третий этап, во время которого за пределами плазменной области образуются области пространственного заряда (ОПЗ). ОПЗ быстро расширяются до тех пор, пока вследствие роста напряженности поля в них не начинается ударная ионизация при $t = t_{av}$ (см.

Рис. 9). Это приводит к сильному замедлению роста напряжения на структуре (см. **Рис. 3**) [20] вплоть до полного восстановления ее блокирующей способности. Однако во время этого заключительного этапа реализуется режим двойной лавинной инжекции, которая может привести к шнурованию тока и разрушению структуры, наблюдавшемуся в ранней экспериментальной работе [7]. Это явление изучалось во многих работах (см. список литературы в обзорах [21,22]), однако в нашем случае оно может обладать рядом нетривиальных особенностей вследствие того, что скорость уменьшения тока в десятки раз превышает обычные значения. Анализу этих особенностей будет посвящена отдельная работа.

### 3.5. Особенности работы фототранзистора и фототиристора

Во время первых двух этапов процессы коммутации фотодиода и фототранзистора практически совпадают. Различия возникают на третьем этапе. В фототранзисторе обычная инжекция дырок из прямо смещенного анодного $n-p^+$-перехода подавляет его восстановление и образование прианодной ОПЗ. Поэтому нарастание напряжения происходит чуть медленнее. На последнем этапе фототранзистор восстанавливает свою блокирующую способность так же, как и обычные биполярные коммутаторы типа GTO или IGBT [23] и может разрушиться по тем же причинам [23,23].

В фототиристоре, кроме того, добавляется обычная инжекция электронов из прямо смещенного катодного $n^+-p$-перехода. Однако она начинает проявляться с задержкой $t_d \sim 10$ ns, по порядку величины равной времени пролета через сильно легированную часть *p*-базы[3] с толщиной $l_p = 8$ μm и концентрацией акцепторов $N_a \sim 2 \cdot 10^{17}$ cm$^{-3}$ [24]. Если $W_{ph} \geq 100$ μJ, то $t_d < t_{0p}$, $p^+-p$-переход не успевает восстановиться и ОПЗ не возникает. В этом простейшем случае, рассмотренном в [14], фототиристор ведет себя как прямо смещенный диод, переходящий со временем в стационарное состояние. При меньших $W_{ph}$ выполняется неравенство $t_d > t_{0p}$, $p^+-p$-переход успевает восстановиться, ОПЗ возникает и начинает расширяться. Но при $W_{ph} \geq 20$ μJ этот процесс подавляется инжекцией электронов, так что примерно через 20 нс напряжение на фототиристоре вновь начинает уменьшаться за счет обычного регенеративного механизма включения. Если же $W_{ph} < 20$ μJ, то это механизм также срабатывает, но только после кратковременной стадии лавинной инжекции. Отметим, что описанный выше немонотонный характер включения фототиристора наблюдался ранее в экспериментах [4,6].

---

[3] Этот слой введен для повышения эффективность шунтировки, которая обеспечивает напряжение переключения более 5 кВ при dU/dt порядка 1 кВ/мкс.